\documentstyle[psfig,12pt,axodraw]{article}
\begin{document}
\baselineskip 18pt
\def\today{\ifcase\month\or
 January\or February\or March\or April\or May\or June\or
 July\or August\or September\or October\or November\or December\fi
 \space\number\day, \number\year}
\def\thebibliography#1{\section*{References\markboth
 {References}{References}}\list
 {[\arabic{enumi}]}{\settowidth\labelwidth{[#1]}
 \leftmargin\labelwidth
 \advance\leftmargin\labelsep
 \usecounter{enumi}}
 \def\newblock{\hskip .11em plus .33em minus .07em}
 \sloppy
 \sfcode`\.=1000\relax}
\let\endthebibliography=\endlist
\begin{titlepage}
\hspace*{10.0cm}ICRR-Report-405-97-28 
   
\hspace*{10.0cm}OCHA-PP-108
\  \
\vskip 0.5 true cm 
\begin{center}
{\large {\bf $CP$ violation in the top-quark system \\ 
  through light squarks }}  
\vskip 2.0 true cm
\renewcommand{\thefootnote}
{\fnsymbol{footnote}}
Mayumi Aoki $^1$ \footnote{Research Fellow of the Japan Society 
for the Promotion of Science.} and Noriyuki Oshimo $^2$ \\
\vskip 0.5 true cm 
{$^1$ \it Graduate School of Humanities and Sciences,  
Ochanomizu University}  \\
{\it Otsuka 2-1-1, Bunkyo-ku, Tokyo 112, Japan}  \\
{$^2$ \it Institute for Cosmic Ray Research, 
University of Tokyo} \\
{\it Midori-cho 3-2-1, Tanashi, Tokyo 188, Japan}  \\
\end{center}

\vskip 4.0 true cm

\centerline{\bf Abstract}
\medskip
 
     We study a decay rate asymmetry of the top quark in 
the supersymmetric standard model, taking into account  
the constraints from the electric dipole moments of the 
neutron and the electron.  
One $CP$-violating phase contained in the mass-squared matrices of   
squarks and sleptons is allowed to have an unsuppressed 
value, even if their masses are of order 100 GeV.   
Sizable $CP$ violation could then occur in the 
production and decay of the top quark through interactions 
with top squarks.  
The asymmetry between the widths for the decays 
$t\rightarrow bW^+$ and $\bar t\rightarrow \bar bW^-$ 
can be of order $10^{-3}$, 
which will possibly be detectable in the near future.   

\medskip


\end{titlepage}

\def\gsim{{\mathop >\limits_\sim}}
\def\lsim{{\mathop <\limits_\sim}}
\def\PL{{1-\gamma_5\over 2}}
\def\PR{{1+\gamma_5\over 2}}
\def\r2{\sqrt 2}
\def\a#1{\alpha_#1}
\def\sw2{\sin^2\theta_W}
\def\tw{\tan\theta_W}
\def\v#1{v_#1}
\def\tb{\tan\beta}
\def\c2b{\cos 2\beta}
\def\w{\omega}
\def\x{\chi}
\def\g{\tilde g}
\def\sq{\tilde q}
\def\su{\tilde u}
\def\sd{\tilde d}
\def\st{\tilde t}
\def\sb{\tilde b}
\def\sl{\tilde l}
\def\m#1{{\tilde m}_#1}
\def\mg{{\tilde m}_g}
\def\mH{m_H}
\def\mgr{m_{3/2}}
\def\mw#1{\tilde m_{\omega #1}}
\def\mx#1{\tilde m_{\chi #1}}
\def\M{\tilde M}

\section{Introduction}  

     Up to now $CP$ violation has only been found in 
the $K^0$-$\bar K^0$ system, 
which is well described by the Kobayashi-Maskawa (KM) 
mechanism of the standard model (SM) \cite{cprev}.  In near-future 
experiments at $B$ factories, $CP$-violating phenomena are  
expected to be observed in the $B^0$-$\bar B^0$ system, possibly providing 
confirmation of the KM mechanism.  
On the other hand, baryon asymmetry of our universe may be an 
outcome of $CP$ violation, though cannot be generated through  
the KM mechanism.  
Furthermore, many extensions of the SM predict new sources 
of $CP$ violation.  Therefore, various possibilities of 
examining $CP$ violation should be studied theoretically 
and experimentally.   

     One of possible reactions for studying $CP$ violation 
is the production and decay of the top quark 
\cite{topcp,topssm,grzadkowski,christova,soni}.  
Although sizable $CP$ violation is not predicted within the 
framework of the SM, 
it could be induced by physics beyond the SM, 
such as supersymmetry \cite{topssm,grzadkowski,christova,soni}.  
In particular, $CP$-violating asymmetries may be observed 
in the angular or energy distributions 
of the particles arising from the top-quark decays.        
These asymmetries have been studied extensively.   

     In this paper we discuss another possible manifestation of $CP$ violation 
in the top-quark decay:  
an asymmetry between the partial decay rates 
for the decays $t\rightarrow bW^+$ and $\bar t\rightarrow \bar bW^-$ 
\begin{equation}
A_{CP}=\frac{\Gamma(t\rightarrow bW^+)-\Gamma(\bar t\rightarrow \bar bW^-)}
       {\Gamma(t\rightarrow bW^+)+\Gamma(\bar t\rightarrow \bar bW^-)}.  
\label{cpasy}
\end{equation} 
We assume the supersymmetric standard model (SSM) 
based on $N=1$ supergravity coupled to grand unified 
theories (GUTs) \cite{ssmrev}, which could contain new 
$CP$-violating phases.    
For this asymmetry to be nonvanishing, in addition to $CP$ violation,  
it is necessary that the top quark has a decay mode  
which can yield a bottom quark and a $W$ boson through a final state 
interaction.  
We show that these two requirements can really be accommodated 
in  the SSM, in spite of severe constraints imposed from the 
electric dipole moments (EDMs) of the neutron and the electron. 
The asymmetry could have a value 
of order $10^{-3}$, which will be detectable 
in the near future at e.g. LHC.  

     In the SSM, $CP$ violation in the production and decay of the top quark 
could be induced through interactions with the squarks of the third generation  
\cite{topssm,grzadkowski,christova,soni}. 
The $CP$-violating effects of these interactions become sizable if relevant 
complex phases are not suppressed and the squark masses 
are of order 100 GeV.  
However, large $CP$-violating phases and small squark masses can easily 
lead to large magnitudes of the EDMs of the neutron and the electron.  
Since experimental upper bounds on these EDMs are fairly small, 
the possible $CP$-violating effects in the top-quark system 
have to be confronted with the constraints from the EDMs.   

     The neutron and electron EDMs   
strongly constrain $CP$ violation in the SSM.   
In general, their experimental bounds are satisfied if   
new $CP$-violating phases are much smaller than unity or 
squarks and sleptons are heavier than 1 TeV \cite{edm}.  
In the ordinary scheme of the SSM, the masses of squarks and sleptons 
are not very different from each other, irrespective of the 
generations that they belong to.  
Also one of the new $CP$-violating phases is independent of the 
generations, and the other new phases are only slightly dependent on them.  
It is not likely that only the squarks of the third generation have 
masses of order 100 GeV and $CP$-violating phases of order unity.   
Therefore, $CP$ violation in the top-quark system is  
severely constrained by the EDMs in the ordinary SSM.     
In the previous works for $CP$ violation in the top-quark system,  
it was assumed explicitly or implicitly that 
squark masses or $CP$-violating phases in the third generation  
are unrelated to those responsible for the EDMs of the 
neutron and the electron.  
The constraints from the EDMs were disregarded.  

     Reexamining the EDMs in the ordinary scheme of the SSM,  
we consider conditions of having   
squark masses of order 100 GeV and an unsuppressed $CP$-violating phase.     
Under these conditions, one of the top squarks can be fairly light, 
while its interactions violate $CP$ invariance maximally.   
It is shown that in a certain parameter region the top quark 
can decay into a top squark $\st$ and a neutralino $\x$, 
being subsequently followed by a scattering which produces  
a bottom quark and a $W$ boson, as shown in Fig. \ref{oneloop}.  
This decay process contributes to the $CP$ asymmetry $A_{CP}$ 
in Eq. (\ref{cpasy}).  

     This paper is organized as follows.  
In sect. 2 we summarize new sources of $CP$ violation in the SSM 
and present the interaction Lagrangians relevant to our discussions.    
In sect. 3 we discuss the SSM parameter values for which 
squarks are light and all the $CP$-violating phases are not suppressed, 
without giving too large values of the EDMs.      
In sect. 4 the decay rate asymmetry for the top quark is calculated.   
Analytic formulae of the asymmetry are given explicitly.  
Conclusions are given in sect. 5.     

\section{$CP$-violating interactions}

     The SSM contains new complex phases for $CP$ violation in addition  to 
the KM phase in the Cabibbo-Kobayashi-Maskawa matrix.  
The parameters which have generally complex values are the 
SU(3), SU(2), and U(1) gaugino masses $\m3$, $\m2$, and $\m1$, 
respectively, the Higgsino mass parameter $\mH$ in the bilinear term 
of the Higgs superfields, and the dimensionless coupling constants 
$A_f$'s and $B$ in the trilinear and bilinear terms of the scalar 
fields, respectively.  We assume unification for the  
gaugino masses at the GUT scale, giving the relation 
$\m3/\a3=\m2/\a2=3\m1/5\a1$.   
Also $A_f$'s are assumed to have the same value of 
order unity at the GUT scale, so that  
their differences at the electroweak scale are 
small and thus can be neglected.    
Then, by redefining particle fields, 
we can take, without loss of generality,  
$\mH$ and $A_f$ as complex, and  
$\tilde m_i$ ($i=1,2,3$) and $B\mH$ as real.   
The complex phases of $\mH$ and $A_f$ are physical 
and become the origins of $CP$ violation,    
which we express as  
\begin{eqnarray}
\mH &=& |\mH|\exp(i\theta), \nonumber \\
A_f &=& A = |A|\exp(i\alpha).   
\label{cpphase}
\end{eqnarray}
These conventions make the vacuum expectation values of the Higgs 
bosons $\v1$ and $\v2$ real.  
 
     The complex parameters lead to complex mass matrices 
for charginos $\w$, neutralinos $\x$, squarks $\sq$, and sleptons $\sl$.  
The mass matrices $M^-$ and $M^0$ for the charginos and the neutralinos 
respectively are given by 
\begin{eqnarray}
    M^- &=& \left(\matrix{\m2 & -g\v1/\r2 \cr
                       -g\v2/\r2 & \mH}        \right), 
\label{chmass} \\
   M^0 &=& \left(\matrix{\m1 &  0  & g'\v1/2 & -g'\v2/2 \cr
                         0  & \m2 & -g\v1/2 &   g\v2/2 \cr
                       g'\v1/2 & -g\v1/2 &   0  & -\mH \cr
                      -g'\v2/2 &  g\v2/2 & -\mH &   0}
           \right). 
\label{nemass}
\end{eqnarray}
These mass matrices are diagonalized to give mass eigenstates as  
\begin{eqnarray}
      C_R^\dagger M^-C_L &=& {\rm diag}(\mw1, \mw2) \quad 
                       (\mw1 <\mw2 ),    \\
N^tM^0N &=& {\rm diag}(\mx1, \mx2, \mx3, \mx4) \quad
                       (\mx1<\mx2<\mx3<\mx4), 
\end{eqnarray}
where $C_R$, $C_L$, and $N$ are unitary matrices.  
The mass of the gluinos $\g$ is expressed by the SU(2) gaugino mass 
as $\m3=(\alpha_3/\alpha_2)\m2$.  

     The mass-squared matrix $M^2_q$ for the squarks corresponding to 
a quark $q$ with mass $m_q$, electric charge $Q_q$, 
and third component of the weak isospin $T_{3q}$ is given by   
\[
    \lefteqn{M^2_q =} \hspace{9cm} 
\]
\[
 \left(\matrix{m_q^2 + \c2b (T_{3q} - Q_q\sw2 )M_Z^2 + \M_{qL}^2 &
                                            m_q (R_q\mH + A^*\mgr) \cr
                   m_q (R_q\mH^* + A\mgr) &
                               m_q^2 +  Q_q\c2b\sw2 M_Z^2 + \M_{qR}^2}
           \right),   
\]
\begin{eqnarray}
   R_q &= & \frac{1}{\tb} \quad (\ T_{3q} = \frac{1}{2}\ ),  
    \quad  \tb \quad (\ T_{3q} = -\frac{1}{2}\ ), 
\label{sqmass} \\
   \tb &= & \frac{\v2}{\v1},  \nonumber 
\end{eqnarray}
where $\M_{qL}^2$ and $\M_{qR}^2$ denote  
the mass-squared parameters for the left-handed squark  
and the right-handed squark, respectively, and 
$\mgr$ is the gravitino mass. 
We have neglected generation mixings.  
The mass eigenstates $\sq_1$ and $\sq_2$ are obtained by diagonalizing 
the mass-squared matrix as  
\begin{equation}
      S_q^\dagger\tilde M_q^2 S_q = {\rm diag}(\M_{q1}^2, \M_{q2}^2) \quad 
                                    (\M_{q1}^2<\M_{q2}^2),
\end{equation} 
where $S_q$ is a unitary matrix.  
The slepton mass-squared 
matrices are obtained by appropriately changing $M_q^2$ in Eq. (\ref{sqmass}).  

     The masses of squarks and sleptons are related to each other.  
At the GUT scale, the mass-squared parameters for all the 
squarks and sleptons are considered to have a common value $\mgr^2$ in  
supersymmetry-breaking terms.  
Then, at the electroweak scale, the values of the mass-squared 
parameters for the squarks of the first two generations 
and all the sleptons are approximately the same, 
\begin{equation} 
\M_{qL}^2 \simeq \M_{qR}^2 \simeq \M_{lL}^2 \simeq \M_{lR}^2 \equiv \M^2.  
\end{equation}
Those for the squarks of the third generation are expressed as 
\begin{eqnarray}
 \M_{tL}^2 &=& \M^2-cm_t^2,  \quad \M_{tR}^2 = \M^2-2cm_t^2,  
    \nonumber \\ 
 \M_{bL}^2 &=& \M^2-cm_t^2,  \quad \M_{bR}^2 = \M^2.    
\end{eqnarray}
The parameters $\M_{tL}^2$, $\M_{tR}^2$, and 
$\M_{bL}^2$ receive quantum corrections through   
Yukawa interactions proportional to the top-quark mass $m_t$, 
with $c = 0.1-1$.  
Under this SSM scheme, all the squark and slepton masses are 
roughly the same, if $\M$ is sufficiently larger than $m_t$.  
For $\M\sim m_t$, the squark masses of the third generation 
become different from the other squark and slepton masses.    

     The complex mass matrices for the  
$R$-odd particles lead 
to $CP$-violating interactions.  
The interaction Lagrangians for these particles relevant to our study are 
given in the followings.  

\noindent
{\it The chargino--neutralino--$W$ boson interactions}:
\begin{eqnarray}
\cal L &= & \frac{g}{\r2}W_\mu^\dagger\bar{\x_j}\gamma^\mu
                  \left(H_{Lji}\PL+H_{Rji}\PR\right)
                           \w_i  +{\rm H.c.},   \\
& & H_{Lji} = \r2 N^*_{2j}C_{L1i}+N^*_{3j}C_{L2i},  \nonumber  \\
& & H_{Rji} = \r2 N_{2j}C_{R1i}-N_{4j}C_{R2i}.    \nonumber  
\end{eqnarray}   
{\it The chargino--quark--squark interactions}:
\begin{eqnarray}
\cal L &=& i\frac{g}{\r2}\sd_k^\dagger \bar{\w_i^c} \left(A_{Li}^k\PL+A_{Ri}^k\PR\right) u  +{\rm H.c.},    \\
& & A_{Li}^k = \r2\left(C_{L1i}S_{d1k}^*
                      -\kappa_d C_{L2i}S_{d2k}^*\right),  \nonumber \\
& & A_{Ri}^k = \r2\kappa_u C_{R2i}S_{d1k}^*,  \nonumber \\ 
\cal L &= & i\frac{g}{\r2}\su_k^\dagger\bar{\w_i}
\left(B_{Li}^k\PL+B_{Ri}^k\PR\right) d  +{\rm H.c},    \\
& & B_{Li}^k = \r2\left(C_{R1i}^*S_{u1k}^*
                     -\kappa_u C_{R2i}^*S_{u2k}^*\right), \nonumber \\ 
& & B_{Ri}^k = \r2\kappa_d C_{L2i}^*S_{u1k}^*,   \nonumber  
\end{eqnarray}
where '$u$' and '$d$' denote up-type and down-type particles, respectively, 
and $\kappa_u$ and $\kappa_d$ are defined by  
\begin{equation}
     \kappa_u = \frac{m_u}{\r2\sin\beta M_W},  \quad
     \kappa_d = \frac{m_d}{\r2\cos\beta M_W}.   
\end{equation} 
{\it The neutralino--quark--squark interactions}:
\begin{eqnarray}
\cal L &=& i\frac{g}{\r2}\su_k^\dagger\bar{\x_j}
\left(F_{Lj}^k\PL+F_{Rj}^k\PR\right) u  +{\rm H.c.},   
\label{neutralinou} \\
& & F_{Lj}^k = \left[(2Q_u - 1)\tw N_{1j} + N_{2j}\right] S_{u1k}^* 
       + \r2\kappa_u N_{4j}S_{u2k}^*, \nonumber \\
& & F_{Rj}^k = 2Q_u\tw N_{1j}^*S_{u2k}^* 
       - \r2 \kappa_u N_{4j}^*S_{u1k}^*, \nonumber \\ 
  \cal L &=& i\frac{g}{\r2}\sd_k^\dagger\bar{\x_j}
\left(G_{Lj}^k\PL+G_{Rj}^k\PR\right) d  +{\rm H.c.},    
\label{neutralinod} \\
& & G_{Lj}^k = \left[(2Q_d + 1)\tw N_{1j} - N_{2j}\right] S_{d1k}^* 
       + \r2\kappa_d N_{3j}S_{d2k}^*, \nonumber \\
& & G_{Rj}^k = 2Q_d\tw N_{1j}^*S_{d2k}^* 
       - \r2 \kappa_d N_{3j}^*S_{d1k}^*.  \nonumber  
\end{eqnarray}
{\it The gluino--quark--squark interactions}:  
\begin{eqnarray}
  \cal L &=& i\r2 g_3\sq_k^\dagger T_3^a\bar{\g}^a
    \left( S_{q1k}^*\PL + S_{q2k}^*\PR\right) q + {\rm H.c.}, 
\label{gluino}
\end{eqnarray}
where $T_3^a$'s represent the generators of the SU(3) group, and 
the color indices are understood.  

\noindent
{\it The squark--squark--$W$ boson interactions}:
\begin{eqnarray}
\cal L &= & i\frac{g}{\r2}D_{kl}\left(\su_k^\dagger\partial^\mu\sd_l
   - \sd_l\partial^\mu\su_k^\dagger\right) W_\mu^\dagger +{\rm H.c.},   \\
& & D_{kl} = S_{u1k}^*S_{d1l}.  \nonumber 
\end{eqnarray}
The interaction Lagrangians for the lepton and slepton sector  
are obtained trivially from these equations.  

     In our scheme, the SSM parameters which determine the interactions 
at the electroweak scale are $\tb$, $A$, $\mH$, $\m2$, $\M$, $\mgr$, and $c$.  
Although these parameters are not all independent of each other,  
they can have various sets of values 
depending on assumptions for underlying GUTs and parameter values.    
For instance, the phase $\theta$ may be related to the phase $\alpha$ 
by taking specific values for them at the GUT scale \cite{garisto}.  
However, if their GUT-scale values are arbitrary, such a  
relation is not obtained.   
Therefore, for simplicity, we take those parameters independent and   
assume only rough constraints coming from theoretical and 
experimental considerations.    

\section{Constraints from EDMs}

     The $CP$-violating interactions of the SSM give rise to 
the EDMs of the neutron and the electron at the one-loop level.  
The exchanged particles in the loop diagrams are  
charginos, neutralinos, or gluinos with squarks or sleptons.  
Among these diagrams, both the neutron and the electron EDMs 
generally receive dominant contributions from the chargino-loop diagrams,   
which are approximately proportional to $\sin\theta$.  
For an unsuppressed value of the $CP$-violating phase $\theta$,    
the experimental upper bounds on the EDMs impose the constraints  
that the squarks and sleptons should be heavier than 1 TeV \cite{edm}.   

     The constraints on the squark and slepton masses are relaxed, 
if the EDMs do not receive contributions from the chargino-loop 
diagrams.  
These chargino contributions become small as the magnitude 
of $\theta$ decreases, irrespective of the value of 
another $CP$-violating phase $\alpha$.  
For a sufficiently small value of $\theta$, therefore, it is expected  
that the squarks and sleptons can have relatively small masses 
even if $\alpha$ is of order unity \cite{aoki}.  

     Assuming $\theta\ll 1$ and $\alpha\sim 1$, 
the SSM parameters are constrained 
by the gluino and neutralino contributions to the neutron EDM, 
and the neutralino contribution to the electron EDM.  
The gluino contribution to the EDM of a quark $q$ is obtained 
from the Lagrangian in Eq. (\ref{gluino}) as  
\begin{equation}
   d_q^G/e = \frac{2\alpha_3}{3\pi}Q_q\sum_{k=1}^2{\rm Im}(S_{q1k}^*S_{q2k})
             \frac{\m3}{\M_{qk}^2}
I\left(\frac{\m3^2}{\M_{qk}^2}, \frac{m_q^2}{\M_{qk}^2}\right).  
\end{equation}
From the Lagrangians in Eqs. (\ref{neutralinou}) and (\ref{neutralinod})  
the neutralino contribution to the EDM of a quark or a lepton $f$ becomes  
\begin{eqnarray}
     d_f^N/e &=& \frac{\alpha_2}{8\pi}Q_f
             \sum_{k=1}^2\sum_{j=1}^4{\rm Im}(K_j^k)
              \frac{\tilde m_{\x j}}{\M_{fk}^2}
I\left(\frac{\tilde m_{\x j}^2}{\M_{fk}^2}, 
                      \frac{m_f^2}{\M_{fk}^2}\right), \\
    K_j^k &=&   F_{Lj}^kF_{Rj}^{k*} \quad (\ T_{3f} = \frac{1}{2}\ ), 
             \quad  G_{Lj}^kG_{Rj}^{k*}  \quad (\ T_{3f} = -\frac{1}{2}\ ).   
\end{eqnarray}
For a quark or a lepton of  
the first generation, the argument $m_f^2/\M_{fk}^2$ 
can be safely neglected.  
The function $I(r,s)$ is written for $s=0$ as 
\begin{equation}
   I(r,0) = \frac{1}{2(1 - r)^2}\left(1 + r + \frac{2r}{1 - r}\log r\right).  
\end{equation}
In terms of the up-quark and down-quark EDMs, the neutron EDM is 
given by $d_n=(4d_d-d_u)/3$ from the nonrelativistic quark model.  
 
     The gluino contribution to the neutron EDM depends on 
the squark and gluino masses.  For the squark masses given, 
the predicted magnitude of the EDM decreases as the gluino mass,  
which is determined by $\m2$, increases.   
The neutralino contributions could also be suppressed by heavy neutralinos, 
whose masses depend on $\m2$, $\mH$, and $\tb$.  
However, the lightest neutralino 
should be lighter than any squarks or sleptons on cosmological grounds.    
All the neutralinos cannot be arbitrarily heavy.  
If $\m2$ is large compared to the squark and slepton masses, 
$|\mH|$ has to be small.   In this case, the lightest neutralino has a mass 
smaller than $|\mH|$ \cite{neutralino}. 
The lighter chargino mass is around $|\mH|$.   
 
\begin{figure}
\caption{The EDMs of the neutron and the electron, being respectively 
    represented by curves (i) and (ii), 
   as a function of $\m2$ for $\alpha=\pi/4$ and $\theta=0$.  
    The other parameters are taken as $\tb=2$, $|\mH|=100$ GeV, 
and $|A|\mgr=\M=200$ GeV. }
\label{edmne}

\vspace{2cm}
%\leavevmode
\psfig{file=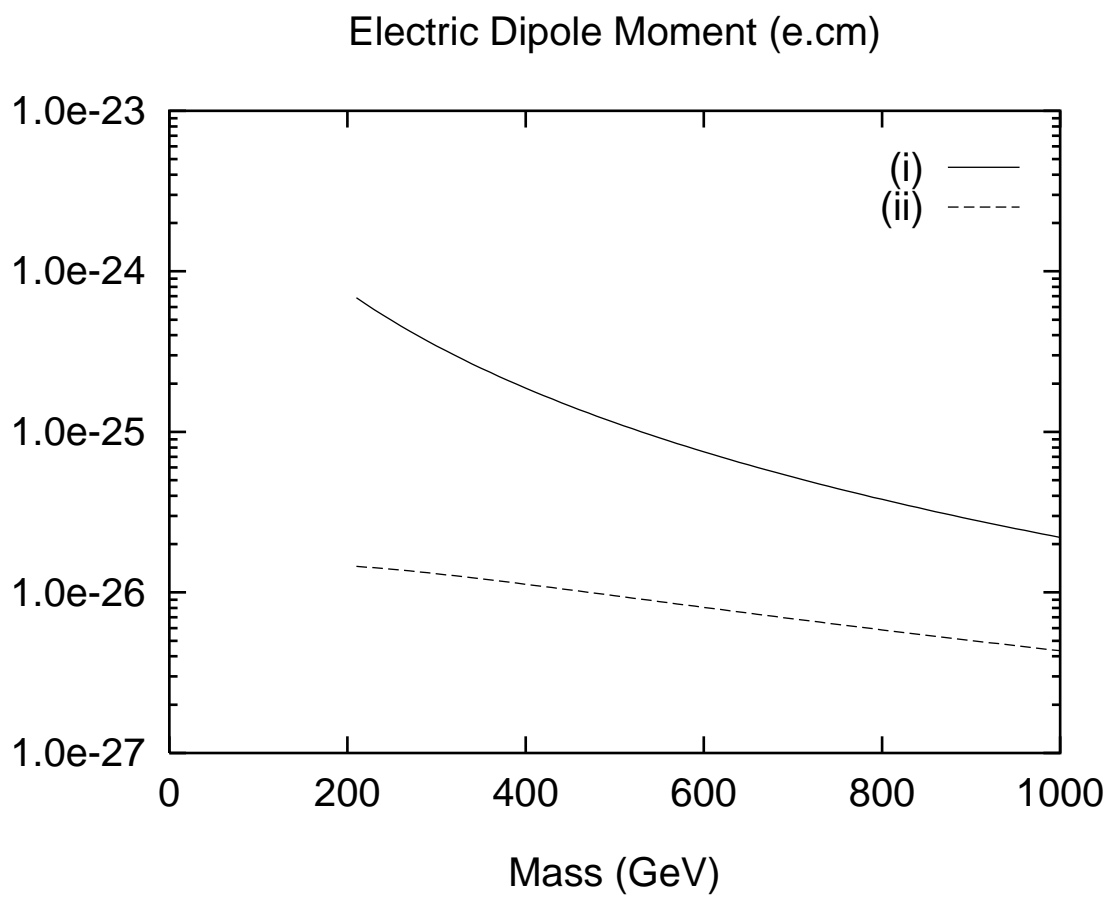}
%%%  options --> height, width, angle
%%%  
\end{figure}

     We show in Fig. \ref{edmne} the absolute value of the EDMs induced 
by light squarks and sleptons with 
$\theta=0$ and $\alpha=\pi/4$ as a function of $\m2$.  
Curves (i) and (ii) respectively represent 
the EDMs of the neutron and the electron. 
The other parameters are taken as $\tb=2$, $|\mH|=100$ GeV, 
and $|A|\mgr=\M=200$ GeV.  
The masses of squarks and sleptons, excluding top and bottom squarks, 
become approximately equal to 200 GeV.  
The signs of the curves are both positive.    
In the mass ranges where curves are not drawn, the lighter chargino is 
lighter than 62 GeV, which is ruled out by experiments \cite{pdg}.  
The experimental bounds  
on the neutron and electron EDMs 
are given by $|d_n|\lsim 1\times 10^{-25}e$cm and 
$|d_e|\lsim 1\times 10^{-26}e$cm, respectively \cite{pdg}.  
The predicted value of the neutron EDM, 
which is dominated by the gluino contribution,   
lies within the experimental bound if $\m2\gsim 500$ GeV.  
The electron EDM is consistent with the experimental bound  
in all the range of $\m2$.  
The EDMs do not vary much with $\tb$ in the range $\tb\sim 1-10$.  

     The above numerical analysis shows that, even if squarks, 
sleptons, one chargino, and some of neutralinos have masses 
of order 100 GeV, the phase $\alpha$ can be of order unity  
without causing inconsistency for the EDMs.  
However, gluinos should be heavier than 1 TeV.  In these 
parameter ranges the mass-squared matrix of top squarks  
becomes a source of $CP$ violation.  
The interactions with the top squarks 
can thus induce $CP$-violating phenomena at the electroweak scale.   
On the other hand, 
the mass-squared matrices of the other squarks and sleptons 
do not lead to sizable $CP$ violation, owing to the 
small imaginary parts of their off-diagonal elements.  

\section{Decay rate asymmetry}

     A nonvanishing value for the decay rate asymmetry $A_{CP}$ 
in Eq. (\ref{cpasy}) could be generated if the decay 
$t\rightarrow\st\x$ is allowed kinematically.  
The produced top squark and neutralino can become a 
bottom quark and a $W$ boson by exchanging charginos or 
bottom squarks as shown in Fig. \ref{oneloop}.  
If $CP$ invariance is violated, the interference of these 
one-loop diagrams and the tree diagram  
makes the partial decay rates 
different between the two decays  
$t\rightarrow bW^+$ and $\bar t\rightarrow\bar bW^-$.  
This difference might also be caused by  
the decays $t\rightarrow \st\g$ and $t\rightarrow \sb\w$.  
However, the gluinos should be much heavier than the 
top quark from the constraint of the neutron EDM.   
Furthermore, the sum of the experimental lower bounds on gluino 
and top-squark masses \cite{pdg} exceeds the top-quark mass.  
The decay $t\rightarrow \st\g$ is not allowed kinematically.  
Since the lighter chargino and the lighter bottom squark are 
respectively heavier than the lightest neutralino and the lighter top squark, 
the parameter values which allow the decay $t\rightarrow \sb\w$ 
are more restricted than for the decay $t\rightarrow \st\x$.  
In fact, it turns out that the former decay is   
forbidden kinematically in most region of parameter space.  

     The decay rate asymmetry is obtained as   
\begin{eqnarray}
A_{CP} &=& \frac{\alpha_2}{2}\left[\biggl\{m_t^2+m_b^2-2M_W^2+
          \frac{(m_t^2-m_b^2)^2}{M_W^2}\biggl\}
    \sqrt{\lambda(m_t^2,M_W^2,m_b^2)}\right]^{-1}
       \left(T^a + T^b\right),   \nonumber  \\
T^a &=& \sum_{i=1}^2\sum_j\sum_k\sum_{n=1}^8{\rm Im}(X_{ijk}^n)
      I_n(\tilde m_{\w i},\tilde m_{\x j},\M_{tk}),   
\label{decayasy} \\
T^b &=& \sum_{l=1}^2\sum_j\sum_k\sum_{n=1}^4{\rm Im}(Y_{ljk}^n)
      J_n(\M_{bl},\tilde m_{\x j},\M_{tk}),  \nonumber
\end{eqnarray}
where coefficients $X_{ijk}^n$, $Y_{ljk}^n$ and functions 
$I_n(\tilde m_{\w i},\tilde m_{\x j},\M_{tk})$, 
$J_n(\M_{bl},\tilde m_{\x j},\M_{tk})$ 
are given in Appendix.  
The dummy suffixes $j$ and $k$ run for kinematically 
allowed pairs of $\st_k$ and $\x_j$.  
The kinematic function $\lambda(a,b,c)$ is defined by 
\begin{equation}
\lambda(a,b,c)=a^2+b^2+c^2-2ab-2bc-2ca.   
\end{equation}
The terms $T^a$ and $T^b$ respectively come from the contributions 
of the diagrams (a) and (b) in Fig. \ref{oneloop}.  
   
\begin{figure}
\caption{The decay rate asymmetry  
   as a function of $\m2$ for $\alpha=\pi/4$ and $\theta=0$.  
   Three curves correspond to three sets of values for $\M$ 
   and $c$ in Table\ \protect\ref{ttasy2}.  
 The other parameters are taken as $\tb=2$, $|\mH|=100$ GeV, 
  and $|A|\mgr=\M$. }
\label{tasy2}

\vspace{2cm}
%\leavevmode
\psfig{file=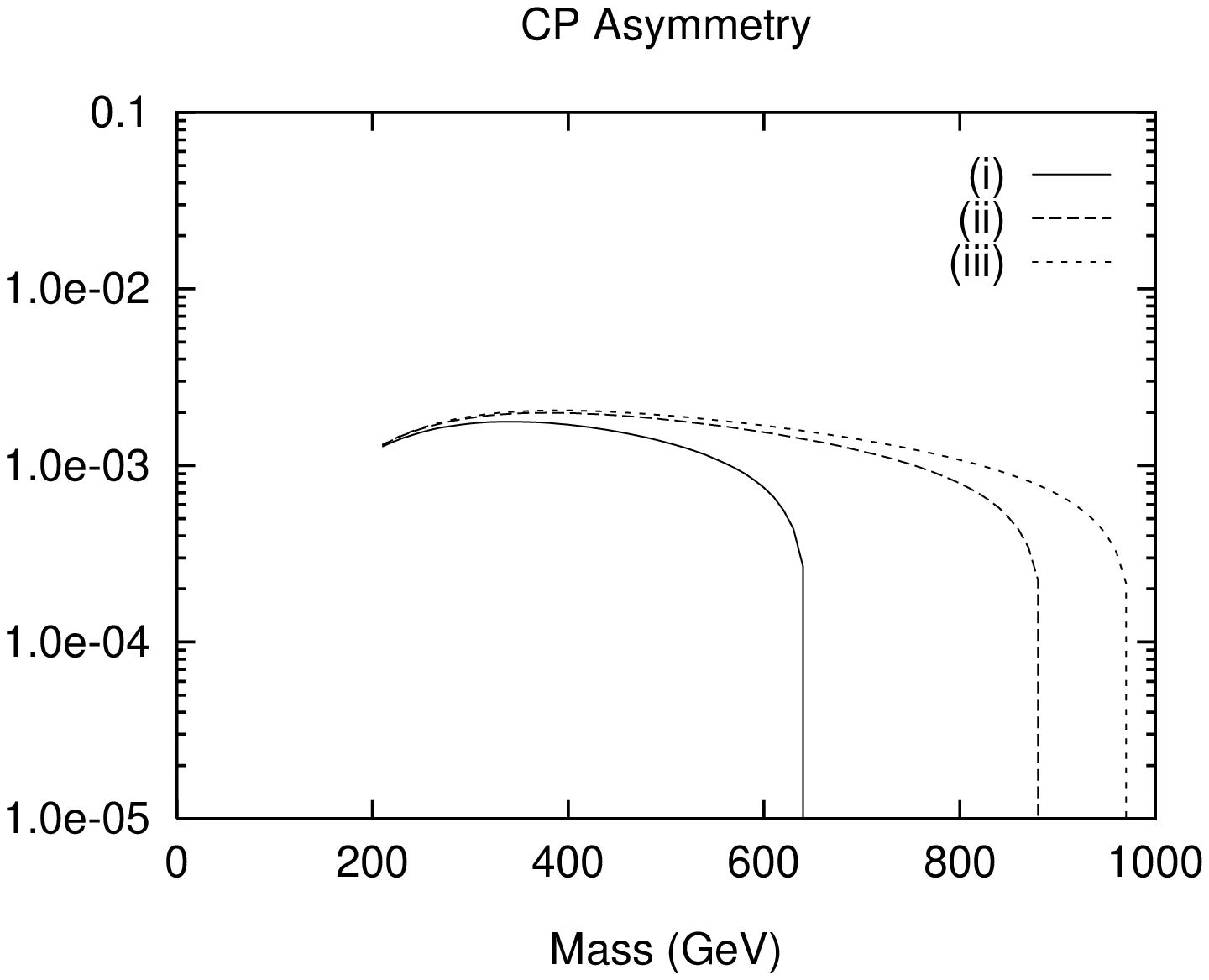}
%%%  options --> height, width, angle
%%%  
\end{figure}

\begin{figure}
\caption{The  decay rate asymmetry 
   as a function of $\m2$ for $\alpha=\pi/4$ and $\theta=0$.  
   Three curves correspond to three sets of values for $\M$ 
   and $c$ in Table\ \protect\ref{ttasy35}.  
 The other parameters are taken as $\tb=35$, $|\mH|=80$ GeV, 
and $|A|\mgr=\M$. }
\label{tasy35}

\vspace{2cm}
%\leavevmode
\psfig{file=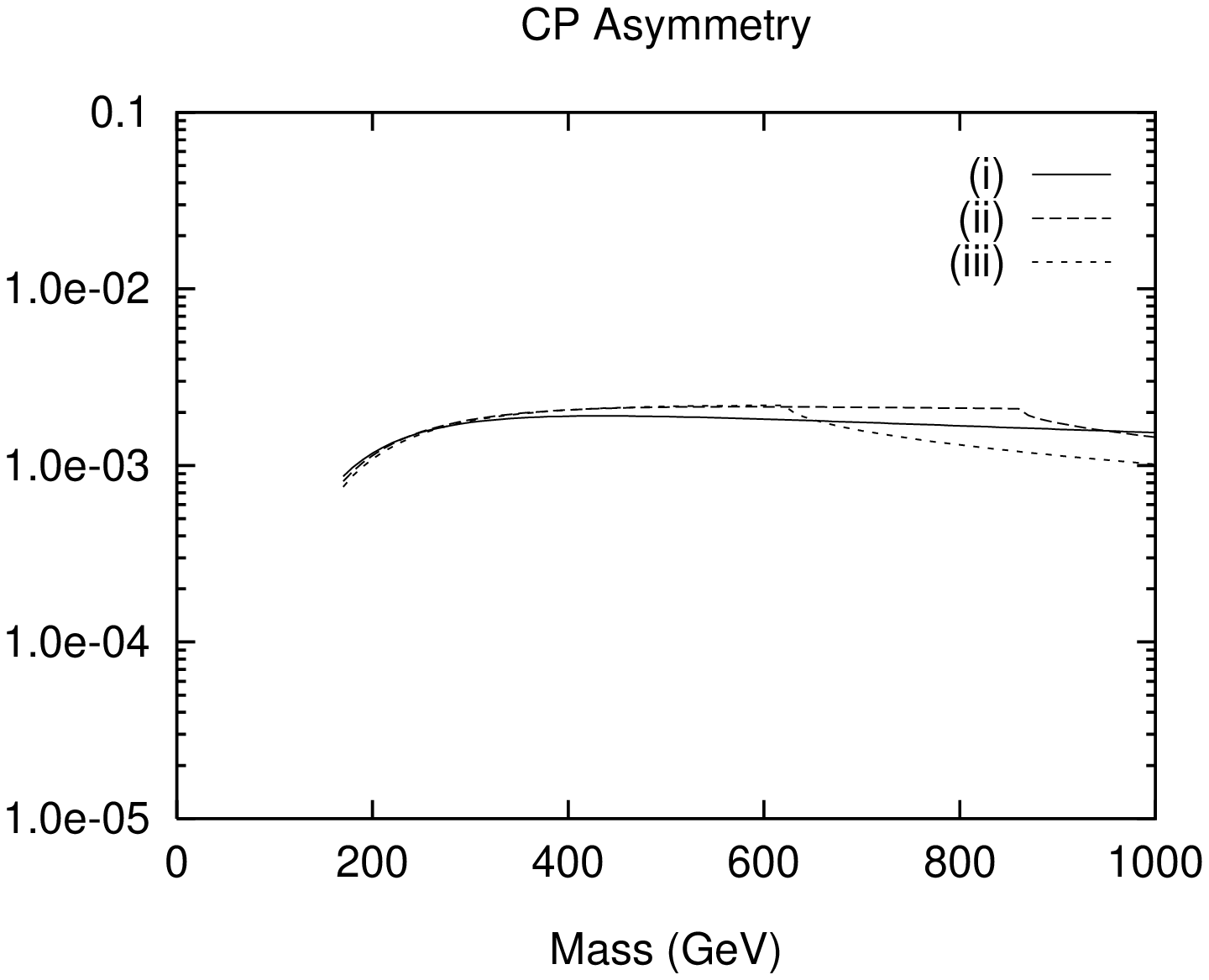}
%%%  options --> height, width, angle
%%%  
\end{figure}

     We show in Figs. \ref{tasy2} and \ref{tasy35} 
the absolute value of the asymmetry $A_{CP}$ as a function 
of $\m2$ for $\theta=0$ and $\alpha=\pi/4$.  
The other parameters are taken so as to satisfy the 
kinematical condition $m_t>\M_{t 1}+\tilde m_{\x 1}$ as well as 
the experimental constraints on the squark and chargino masses \cite{pdg}.     
The lighter top squark, which is the lightest among the squarks 
and sleptons, should be heavier than the lightest neutralino.   
The parameter values are taken in Fig. \ref{tasy2} as   
$\tb=2$, $|\mH|=100$ GeV, and the values of $\M$ and $c$  
given in Table \ref{ttasy2}, and in Fig. \ref{tasy35} as   
$\tb=35$, $|\mH|=80$ GeV, and the values of $\M$ and $c$  
given in Table \ref{ttasy35}.  
We take $|A|\mgr=\M$.  
The top-quark mass is taken to be 180 GeV.  
In Tables \ref{ttasy2} and \ref{ttasy35}, resultant values 
of the lighter top-squark mass are also given.  
In the smaller mass ranges where curves are not drawn, the lighter 
chargino mass is smaller than 62 GeV.  
For large values of $\m2$ in Fig. \ref{tasy2}, the decay 
$t\rightarrow \st\x$ is not allowed kinematically. 

     The asymmetry $A_{CP}$ is of order $10^{-3}$  
in a wide region of the parameter space where the decay 
$t\rightarrow \st\x$ is allowed kinematically. 
For $\m2\gsim 500$ GeV, the predicted value of 
the neutron EDM does not contradict its experimental bound.   
The electron EDM is also predicted to be within its 
experimental bound for all the range of $\m2$.  
This amount of $A_{CP}$ necessitates $t\bar t$ pairs 
of order $10^{6}$ for detection.  
Since $t\bar t$ pairs are expected to be produced at a rate of 
order $10^7$ at LHC, the asymmetry will be detectable in the near future.      
The signs of all the curves are negative.  
The spikes of curves (ii) and (iii) in Fig. \ref{tasy35} 
arise from the opening of the threshold for the decay 
$t\rightarrow \st_1\x_2$.  
For the smaller mass ranges of these curves and all the ranges 
of the other curves shown, allowed two-body decays of the top quark 
are only $t\rightarrow bW$ and $t\rightarrow \st_1\x_1$.  

     The interactions which induce a nonvanishing value of $A_{CP}$ 
also yield a rate difference between the two decays 
$t\rightarrow \st\x$ and $\bar t\rightarrow \st^*\x$.  
These decay widths satisfy the relation 
\begin{equation}
\Gamma(t\rightarrow bW^+)-\Gamma(\bar t\rightarrow \bar bW^-)=
-\left\{\Gamma(t\rightarrow \st\x)-\Gamma(\bar t\rightarrow \st^*\x)\right\}.  
\label{cpt}
\end{equation}  
Therefore, the total width of the top quark is the same as that 
of the anti-top quark, as required by $CPT$ invariance.  

     We briefly comment on some problems encountered when 
rates of the decays 
$t\rightarrow bW^{+}$ and $\bar t\rightarrow \bar bW^{-}$  
are measured in experiments.   
If these two decays occur at different rates, so do the two decays 
$t\rightarrow \st\x$ and $\bar t\rightarrow \st^*\x$.    
In most of the parameter region where the asymmetry $A_{CP}$ is sizable,  
the top squark decays sequentially as $\st\rightarrow b\w$, 
$\w\rightarrow W^*\x$, where $W^*$ denotes the virtual state of $W$.   
Consequently, the decay $t\rightarrow \st\x$ results in a  
final state which contains the same 
visible particles as the decay $t\rightarrow bW$.     
Owing to the relation in Eq. (\ref{cpt}),  
a naive asymmetry between specific particles and their antiparticles 
in the final states of $t\rightarrow bW^+$ and 
$\bar t\rightarrow \bar bW^-$ compensates that in the final 
states of $t\rightarrow \st\x$ and $\bar t\rightarrow \st^*\x$.  
In order to measure $A_{CP}$, therefore, some way is needed to distinguish  
$t\rightarrow bW$ from $t\rightarrow \st\x$.  
In addition, if the top squark has such a small mass, 
pairs of $\st\st^*$ are directly produced at a rate larger than 
the pair production of $t\bar t$.  
Since the decays $t\rightarrow bW$ and $\st\rightarrow b\w$  
yield the same visible particles, 
we also need a way to distinguish between them.   
These distinctions should be made by detailed analyses of energy spectra 
of the particles in the final state.  

\begin{figure}
\caption{The branching ratio of $t\rightarrow \st_1\x_1$   
   as a function of $\m2$ for the parameter values 
  in Fig.\ \protect\ref{tasy2}.}  
\label{bra2}

\vspace{2cm}
%\leavevmode
\psfig{file=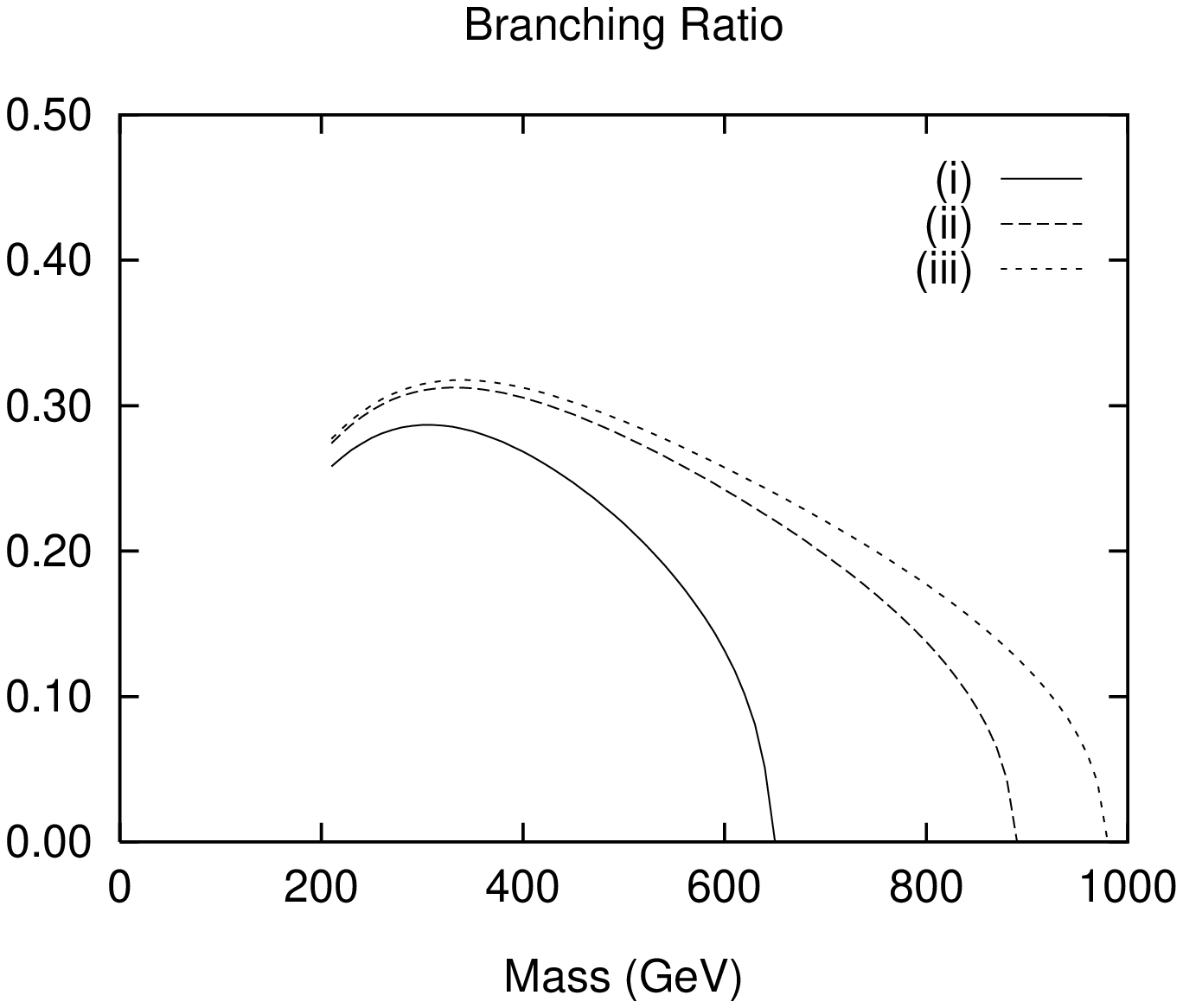}
%%%  options --> height, width, angle
%%%  
\end{figure}

\begin{figure}
\caption{The branching ratio of $t\rightarrow \st_1\x_1$   
   as a function of $\m2$ for the parameter values 
   in Fig.\ \protect\ref{tasy35}.}  
\label{bra35}

\vspace{2cm}
%\leavevmode
\psfig{file=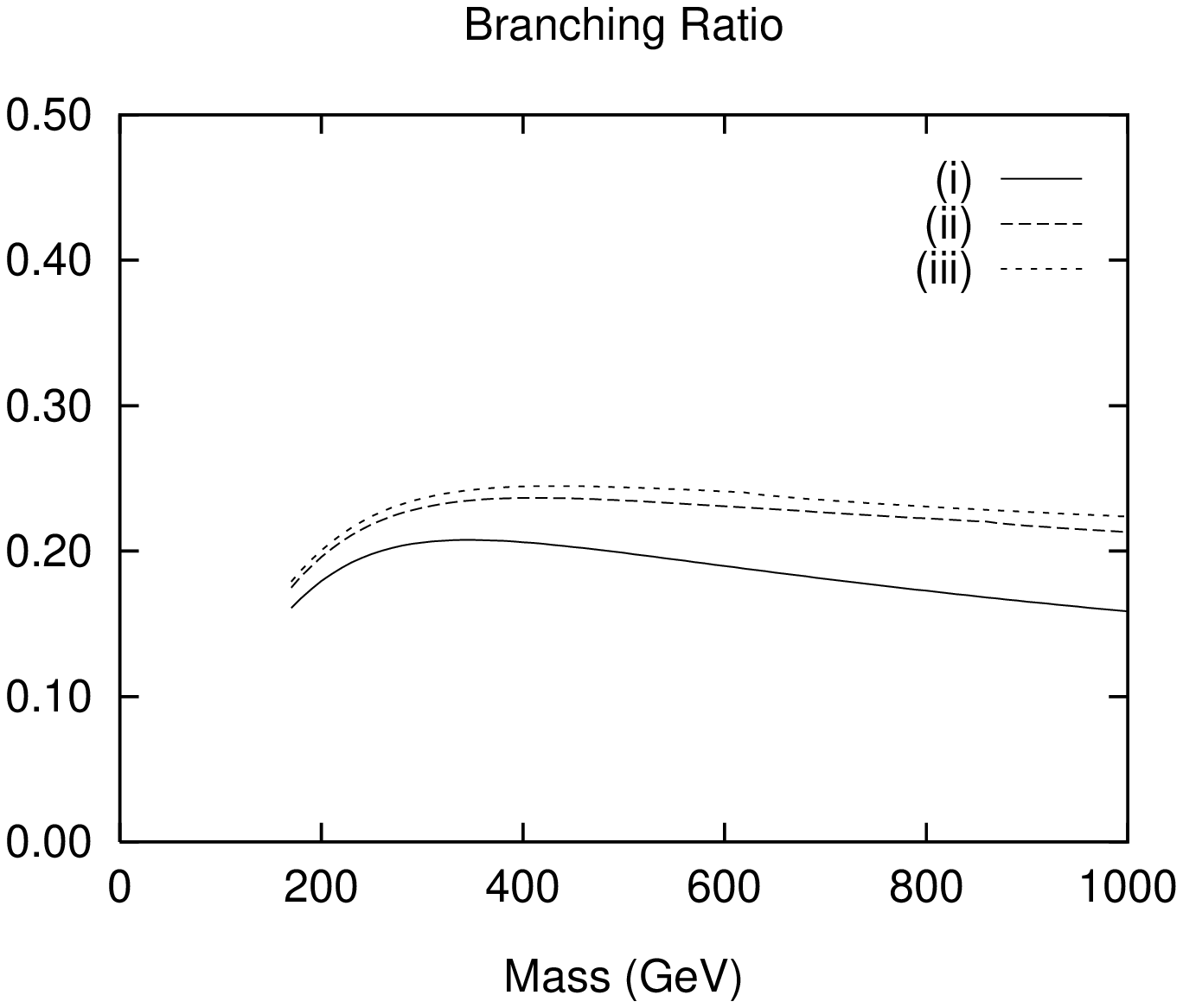}
%%%  options --> height, width, angle
%%%  
\end{figure}

     The non-standard decay $t\rightarrow \st\x$, required   
by a nonvanishing value of the asymmetry $A_{CP}$,  
may be detected at Tevatron, if it has a large branching ratio.         
In Figs. \ref{bra2} and \ref{bra35} the branching ratio of 
$t\rightarrow \st_1\x_1$ is shown for the same parameter 
values as in Figs. \ref{tasy2} and \ref{tasy35}.  
The branching ratio becomes around 0.2.   
As discussed above, 
the final states of the decays $t\rightarrow \st\x$, $t\rightarrow bW$,   
and $\st\rightarrow b\w$ arising from a $\st\st^*$ pair production 
have the same visible particles.  
Energy spectra of these particles have to be examined in detail   
to find the non-standard decay mode.  
In fact, it is not ruled out \cite{sender} that this decay may have   
escaped detection, even if its branching ratio is comparable to that of the 
standard decay $t\rightarrow bW$.    

     It may happen by some reasons that  
the squarks of the third generation are of order 100 GeV 
while the other squarks and sleptons are heavier than 1 TeV.   
Then, the EDMs of the neutron and the electron do not impose 
constraints on the $CP$-violating phases $\theta$ and $\alpha$ 
or the mass parameters $\m2$ and $|\mH|$.  
We have computed the $CP$ asymmetry $A_{CP}$ without imposing the 
constraints of the EDMs, searching for a parameter region 
which gives a large value for $A_{CP}$.  
However, taking into account experimental lower bounds on 
the gluino and top-squark masses,  
the decay $t\rightarrow\st\g$, which might lead to 
a large asymmetry \cite{grzadkowski}, is not allowed kinematically. 
The decay $t\rightarrow\sb\w$ is also not allowed  
in most region of parameter space.  
The asymmetry $A_{CP}$ 
induced by the decay $t\rightarrow \st\x$ is at most around    
$3\times 10^{-3}$, and not significantly large compared to 
the asymmetry obtained under the constraints of the EDMs.  
This numerical result agrees with Ref. \cite{soni} but is much smaller than  
the result of Ref. \cite{christova}.

\section{Conclusions} 

     We have considered a possibility that $CP$ violation 
is observed at the electroweak energy scale within the framework of the SSM.  
In this model, the EDMs of the neutron and the electron severely 
constrain $CP$-violating phases and masses of $R$-odd particles.  
However, we showed that there is a parameter region where
squarks, sleptons, one chargino, and some of neutralinos have masses 
of order 100 GeV and one $CP$-violating phase $\alpha$ is of order unity.   
In this region, the mass-squared matrix of top squarks  
becomes a new source of $CP$ violation.  
This source can affect the production and decay of the top quark, 
which is and will be studied extensively in experiments.  

     Among phenomena possibly induced by the new source 
of $CP$ violation, we studied a decay rate asymmetry of the top quark.  
If the two-body decay $t\rightarrow \st\x$ is allowed,   
the decay widths for $t\rightarrow bW^+$ and 
$\bar t\rightarrow \bar bW^-$ could be different from each other.  
It was shown that this asymmetry is of order $10^{-3}$ in a wide region of 
the parameter space where the decay $t\rightarrow \st\x$ occurs.   
Since  $10^7$ pairs of top quarks are expected to be produced at LHC, 
such an amount of asymmetry will be detectable.  
The decay rate asymmetry necessitates the two-body decay mode 
of the top quark different from the standard decay $t\rightarrow bW$, 
so that the parameter region may be probed at Tevatron.  
If further experimental analyses find the 
non-standard decay of the top quark, the decay rate asymmetry will 
be worth measuring at LHC.  

     The new source of $CP$ violation could affect 
the angular distributions of the particles produced by the top-quark decay.  
Although the decay rate asymmetry 
becomes nonvanishing only in a parameter region 
restricted by the mass relation $m_t>\M_{t1}+\tilde m_{\x 1}$, 
the $CP$ asymmetries of the angular distributions can occur  
in wider parameter region.  
It would be interesting to study these asymmetries   
taking into consideration the results of this paper.  
A light top squark and an unsuppressed $CP$-violating phase 
could also affect the $B$-meson system.   

     The baryon asymmetry of our universe may have been generated 
at the electroweak phase transition.  This scenario could be 
realized in the SSM, if there exists a light top squark with  
$CP$-violating interactions \cite{aoki}.  
This possibility will become more plausible, if $CP$ violation 
in the top-quark system is found.  

\medskip 

\section*{Acknowledgments}

     We thank J. Arafune, J. Kamoshita, A. Sugamoto, 
I. Watanabe, and P. Zerwas for discussions.  
The work of M.A. is supported in part by the Grant-in-Aid for Scientific 
Research from the Ministry of Education, Science and Culture, Japan.  
This work is supported in part by 
the Grant-in-Aid for Scientific Research (No. 08044089) 
from the Ministry of Education, Science and Culture, Japan.  

\section*{Appendix}

\def\mxx{\tilde m_{\chi}}
\def\mst{\tilde M_t}
\def\mtt{m_t}
\def\mw{M_W}
\def\mbb{m_b}
\def\msb{\tilde M_b}
\def\mch{\tilde m_{\omega}}
\def\mx2{\tilde m_{\chi}^2}
\def\mst2{\tilde M_t^2}
\def\mt2{m_t^2}
\def\mw2{M_W^2}
\def\mb2{m_b^2}
\def\msb2{\tilde M_b^2}
\def\mc2{\tilde m_{\omega}^2}
\def\laba{\log \Biggl| \frac{A+B}{A-B}\Biggr|}
\def\lab2{\log \Biggl| \frac{A'+B}{A'-B}\Biggr|}
\def\be{\begin{equation}}
\def\ee{\end{equation}}
\def\bea{\begin{eqnarray}}
\def\eea{\end{eqnarray}}
\def\bean{\begin{eqnarray*}}
\def\eean{\end{eqnarray*}}
\def\nn{\nonumber}

     The coefficients $X_{ijk}^n$ and the functions  
$I_n(\tilde m_\w,\tilde m_\x,\M_t)$ in Eq. (\ref{decayasy}) are defined by 
\begin{eqnarray} 
 & & X_{ijk}^1 = B_{Li}^{k*}F_{Lj}^k H_{Lji}^*,  
       \quad X_{ijk}^2 = B_{Li}^{k*}F_{Lj}^k H_{Rji}^*, 
 \quad X_{ijk}^3 = B_{Li}^{k*}F_{Rj}^k H_{Lji}^*,    \nonumber \\
& & X_{ijk}^4 = B_{Li}^{k*}F_{Rj}^k H_{Rji}^*,  
       \quad X_{ijk}^5 = B_{Ri}^{k*}F_{Lj}^k H_{Lji}^*, 
 \quad X_{ijk}^6 = B_{Ri}^{k*}F_{Lj}^k H_{Rji}^*,    \nonumber \\
& & X_{ijk}^7 = B_{Ri}^{k*}F_{Rj}^k H_{Lji}^*,  
 \quad X_{ijk}^8 = B_{Ri}^{k*}F_{Rj}^k H_{Rji}^*,    
\end{eqnarray}
\bean
 I_1 &=& - \frac{1}{2} \mch \mxx 
\left\{\mt2+\mb2-2\mw2+ \frac{1}{\mw2}(\mt2-\mb2)^2 \right\} \laba, \nn \\
 I_2 &=& - \left\{\mt2(\mt2-\mb2+3\mw2-\mst2+2\mc2-\mx2)+
(\mb2-3\mw2)(\mst2-\mx2)\right\} B \nn \\
&&+\frac{1}{2}\Biggl(\mt2(2\mb2-2\mst2+\mc2)
-\mb2(2\mst2-\mx2) +2(\mst2-\mc2)(\mst2-\mx2)  \nn \\ 
&&+\frac{1}{\mw2}\Biggl[\mt2\left\{\mt2\mc2-\mb2(\mc2+\mx2)   
-(\mst2-\mc2)(\mc2-\mx2)\right\}   \nn \\
&&  +\mb2\left\{\mb2\mx2+(\mst2-\mx2)(\mc2-\mx2)\right\} \Biggr]
 \Biggr) \laba,   \nn \\
 I_3 &=& - \mtt\mch \Biggl[ 2(\mt2-\mb2-2\mw2) B 
+\frac{1}{2}\Biggl\{2\mt2-\mb2-2\mw2-\mst2+2\mc2-\mx2  \nn \\
&&  -\frac{1}{\mw2}(\mt2-\mb2)(\mc2-\mx2)\Biggr\}\laba \Biggr], \nn \\
 I_4 &=& -3 \mtt\mxx  
\left\{2\mw2 B -\frac{1}{2}(\mb2-\mst2+\mc2)\laba \right\}, \nn \\
 I_5 &=& -\mbb \mxx 
\Biggl[ 2\left(\mt2-\mb2+\mw2\right)B 
-\frac{1}{2}\Biggl\{\mt2-2\mb2+2\mw2+\mst2+\mc2-2\mx2  \nn \\
&&  +\frac{1}{\mw2} (\mt2-\mb2)(\mc2-\mx2)\Biggr\} \laba \Biggr], \nn \\
 I_6 &=& \frac{3}{2} \mbb \mch (\mt2-\mst2+\mx2)\laba, \nn \\
 I_7 &=& - \mtt\mbb  
\Biggl[\Biggl\{\mt2-\mb2-3\mw2-\mst2-2\mc2+3\mx2  \nn \\
&&+ \frac{1}{\mt2}(\mb2-\mw2)(\mst2-\mx2)\Biggr\} B \nn \\
&&-\frac{1}{2}\left\{2\mw2-\mc2-\mx2
-\frac{1}{\mw2}(\mc2-\mx2)^2\right\}\laba \Biggr], \nn \\
 I_8 &=& 3\mtt\mbb\mch\mxx\laba,  \nn
\eean
where
\bean
A &=& \frac{(\mt2 +\mw2 -\mb2)(\mt2-\mst2+\mx2)}{4\mw2\mt2}-
        \frac{\mw2+\mx2-\mc2}{2\mw2}, \nn \\
B &=& \frac{1}{4\mt2\mw2}\sqrt{\lambda(\mt2,\mw2,\mb2) 
\lambda(\mt2,\mst2,\mx2)}.  
\eean

    The coefficients $Y_{ljk}^n$ and the functions  
$J_n(\M_b,\tilde m_\x,\M_t)$ are defined by 
\begin{eqnarray}
& & Y_{ljk}^1 = D_{kl}^*F_{Lj}^k G_{Lj}^{l*}, 
     \quad  Y_{ljk}^2 = D_{kl}^*F_{Lj}^k G_{Rj}^{l*}, 
     \quad  Y_{ljk}^3 = D_{kl}^*F_{Rj}^k G_{Lj}^{l*},    \nonumber \\
& & Y_{ljk}^4 = D_{kl}^*F_{Rj}^k G_{Rj}^{l*},   
\end{eqnarray}
\bean
 J_1 &=& -\left\{\mt2(\mt2-\mb2+\mw2+\mst2-2\msb2+\mx2)
+(\mb2-3\mw2)(\mst2-\mx2)\right\} B \nn \\
&& +\frac{1}{2}\Biggl[\mt2(\msb2+\mx2)+\mb2(\mst2+\mx2)
-2(\mst2-\mx2)(\msb2-\mx2)-2\mw2\mx2   \nn \\
&& +\frac{1}{\mw2}\left\{\mt2(\msb2-\mx2)
-\mb2(\mst2-\mx2)\right\}(\mst2-\msb2)\Biggr]\lab2, \nn  \\
 J_2 &=& -\mbb\mxx 
\Biggl[2(\mt2-\mb2+\mw2)B 
- \frac{1}{2}\Biggl\{\mt2+\mb2-\mw2+\mst2+\msb2-2\mx2  \nn \\
&&  -\frac{1}{\mw2}(\mt2-\mb2)(\mst2-\msb2)\Biggr\} 
\lab2 \Biggr], \nn \\
 J_3 &=& \frac{\mtt}{\mbb}J_2, \nn \\
 J_4 &=& -\mtt\mbb  
\Biggl[\Biggl\{\mt2-\mb2-\mw2-3\mst2+2\msb2+\mx2 \nn \\ 
&& +\frac{1}{\mt2}(\mb2-\mw2)(\mst2-\mx2)\Biggr\}B 
- \frac{1}{2}\Biggl\{\mt2+\mb2-\mst2-\msb2-2\mx2  \nn \\ 
&&  -\frac{1}{\mw2}(\mt2-\mb2-\mst2+\msb2)(\mst2-\msb2)\Biggr\}
\lab2 \Biggr], \nn \\
\eean
where
\bean
A' &=& \frac{(\mt2 +\mw2 -\mb2)(\mt2+\mst2-\mx2)}{4\mw2\mt2}-
        \frac{\mw2+\mst2-\msb2}{2\mw2}. \nn 
\eean

\newpage

\newpage 

\begin{table}
\caption{The values of $\M$ and $c$ for curves 
            (i)--(iii) in Fig.\ \protect\ref{tasy2}. 
        The lighter top-squark mass is also given.}
\label{ttasy2}

\vspace{1cm}
\begin{center}
\begin{tabular}{cccc}
    & $\M$ (GeV) & c & $\M_{t1}$ (GeV) \\
\hline 
 (i)   & 180 & 0.3 &  97   \\
 (ii)  & 200 & 0.4 &  92   \\
 (iii) & 220 & 0.5 &  91   \\
\end{tabular} 
\end{center}
\end{table}

\begin{table}
\caption{The values of $\M$ and $c$ for curves 
            (i)--(iii) in Fig.\ \protect\ref{tasy35}. 
        The lighter top-squark mass is also given.}
\label{ttasy35}

\vspace{1cm}
\begin{center}
\begin{tabular}{cccc}
    & $\M$ (GeV) & c & $\M_{t1}$ (GeV) \\
\hline 
 (i)   & 180 & 0.4 & 101   \\
 (ii)  & 200 & 0.5 &  95   \\
 (iii) & 220 & 0.6 &  93   \\
\end{tabular} 
\end{center}
\end{table}

\newpage   

\begin{figure}
\caption{The one-loop diagrams for the top-quark decay 
producing a bottom quark and a $W$ boson,     
where $\st$, $\sb$, $\w$, and $\x$ denote respectively 
the top squark, bottom squark, chargino, and neutralino.  
  }
\label{oneloop}

\vspace{2cm}

%
%\unitlength=1mm
%\SetScale{2}
\begin{center}\begin{picture}(400,100)(0,0)
\Line(23,100)(70,100)
\Line (70,100)(130,130)
\DashLine (70,100)(130,70){5}
\Line(130,130)(130,70)
\Photon(130,130)(180,130){5}{3}
\Line(130,70)(180,70)
\Text(17,100)[]{$t$}
\Text(100,125)[]{$\chi$}
\Text(100,75)[]{$\tilde t$}
\Text(140,100)[]{$\omega$}
\Text(188,130)[]{$W$}
\Text(188,70)[]{$b$}
\Text(105,50)[]{(a)}
%%%%
\Line(223,100)(270,100)
\DashLine (270,100)(330,130){5}
\Line (270,100)(330,70)
\DashLine(330,130)(330,70){5}
\Photon(330,130)(380,130){5}{3}
\Line(330,70)(380,70)
\Text(217,100)[]{$t$}
\Text(300,125)[]{$\tilde t$}
\Text(300,75)[]{$\chi$}
\Text(340,100)[]{$\tilde b$}
\Text(388,130)[]{$W$}
\Text(388,70)[]{$b$}
\Text(305,50)[]{(b)}
\end{picture}
\end{center}

\end{figure}
\pagebreak

\vspace{2cm}

\pagebreak

\vspace{2cm}

\pagebreak

\vspace{2cm}

\pagebreak

\vspace{2cm}

\pagebreak

\vspace{2cm}


\begin{thebibliography}{99}
%
\bibitem{cprev}
           For general reviews, see e.g. 
           J.F. Donoghue, B.R. Holstein, and G. Valencia, 
           Int. J. Mod. Phys. A2 (1987) 319; \\  
           W. Grimus, Fortschr. Phys. 36 (1988) 201.  
\bibitem{topcp} 
  G. Eilam, J.L. Hewett, and A. Soni, Phys. Rev. Lett. 67 (1991) 1979; \\ 
  D. Atwood and A. Soni, Phys. Rev. D45 (1992) 2405;  \\
  C. R. Schmidt and M. E. Peskin, Phys. Rev. Lett. 69 (1992) 410; \\
  B. Grzadkowski and J.F. Gunion, Phys. Lett. B287 (1992) 237; \\
  D. Atwood, G. Eilam, A. Soni, R.R. Mendel, and R. Migneron, 
            Phys. Rev. Lett. 70 (1993) 1364; \\
  D. Chang, W.-Y. Keung, and I. Phillips, Nucl. Phys. B408 (1993) 286; \\
  W. Bernreuther and A. Brandenburg, Phys. Lett. B314 (1993) 104; \\
  F. Cuypers and S. D. Rindani, Phys. Lett. B343 (1995) 333; \\
  P. Poulose and S.D. Rindani, Phys. Lett. B349 (1995) 379; \\
  B. Grzadkowski and Z. Hioki, Phys. Lett. B391 (1997) 172; \\
  T. Hasuike, T. Hattori, T. Hayashi, and S. Wakaizumi, 
       Z. Phys. C76 (1997) 127.  
\bibitem{topssm} 
   C. R. Schmidt, Phys. Lett. B 293 (1992) 111; \\
   E. Christova and M. Fabbrichesi, Phys. Lett. B315 (1993) 338; \\
   B. Grzadkowski and W.-Y. Keung, Phys. Lett. B316 (1993) 137;  \\
   A. Bartl, E. Christova, and W. Majerotto, Nucl. Phys. B460 (1996) 235; \\
   A. Bartl, E. Christova, T. Gajdosik, and W. Majerotto, Nucl. Phys. B507 
    (1997) 35.  
\bibitem{grzadkowski}
  B. Grzadkowski and W.-Y. Keung, Phys. Lett. B319 (1993) 526.  
\bibitem{christova}
  E. Christova and M. Fabbrichesi, Phys. Lett. B320 (1994) 299. 
\bibitem{soni}
   S. Bar-Shalom, D. Atwood, and A. Soni, UCRHEP-T201. 
\bibitem{ssmrev} 
      For reviews, see e.g. 
           H.P. Nilles, Phys. Rep. 110 (1984) 1;  \\
           P. Nath, R. Arnowitt, and A.H. Chamseddine, {\it Applied 
           N=1 Supergravity} 
           (World Scientific, Singapore, 1984); \\
           H.E. Haber and G.L. Kane, Phys. Rep. 117 (1985) 75.   
\bibitem{edm} 
    Y. Kizukuri and N. Oshimo, Phys. Rev. D45 (1992) 1806; D46 (1992) 3025.  
\bibitem{garisto}
   R. Garisto and J.D. Wells, Phys. Rev. D55 (1997) 1611.  
\bibitem{aoki}
  M. Aoki, A. Sugamoto, and N. Oshimo, Prog. Theor. Phys. 98 (1997) 1325.  
\bibitem{neutralino}
A. Bartl, H. Fraas, W. Majerotto, and N. Oshimo, Phys. Rev. D40 (1989) 1594.
\bibitem{pdg}
    Particle Data Group, Phys. Rev. D54 (1996) 1 (updated, 1997).   
\bibitem{sender}
  J. Sender, Phys. Rev. D54 (1996) 3271.  
  
\end{thebibliography}
\end{document}